# Digging into Primary Financial Market: Challenges and Opportunities of Adopting Blockchain

Ji Liu [a,b 1*], Zheng Xu [c], Yanmei Zhang [d], Wei Dai [e], Hao Wu [a], and Shiping Chen [a,b]

[a] School of Electrical & Information Engineering, the University of Sydeny, Australia

[b] CSIRO DATA61, Australia

[c] Chinese Academy of Fiscal Sciences, China

[d] Information School, Central University of Finance and Economics, China

[e] Finance School, Central University of Finance and Economics, China

*Abstract-* Since the emergence of blockchain technology, its application in the financial market has always been an area of focus and exploration by all parties. With the characteristics of anonymity, trust, tamper-proof, etc., blockchain technology can effectively solve some problems faced by the financial market, such as trust issues and information asymmetry issues. To deeply understand the application scenarios of blockchain in the financial market, the issue of securities issuance and trading in the primary market is a problem that must be studied clearly. We conducted an empirical study to investigate the main difficulties faced by primary market participants in their business practices and the potential challenges of the deepening application of blockchain technology in the primary market. We adopted a hybrid method combining interviews (qualitative methods) and surveys (quantitative methods) to conduct this research in two stages. In the first stage, we interview 15 major primary market participants with different backgrounds and expertise. In the second phase, we conducted a verification survey of 54 primary market practitioners to confirm various insights from the interviews, including challenges and desired improvements. Our interviews and survey results revealed several significant challenges facing blockchain applications in the primary market: complex due diligence, mismatch, and difficult monitoring. On this basis, we believe that our future research can focus on some aspects of these challenges.

## I. Introduction

The financial market can divide into primary market and secondary market [1]. The primary market is a market where companies conduct transactions before IPO. Stock and bonds have not undergone a standardized review process before going to the public. Thus, there are many issues such as trust, authenticity, privacy, and etc. in the market. The secondary market is a market for stock/bond trading transactions that have been standardized and reviewed by regulation. There are four parts in the transaction chain of the primary market: raising funds, investing assets, asset monitor, and existence. There are financing *f*appcompanies, investment institutions, brokers, valuation firms, law firms, accounting firms, etc., participating in the primary market. Each participant has a unique purpose and role. These participants help the market to maintain stability.

The primary financial market is enormous. The 'McKinsey Global Private Markets Review 2020' [2] shows that total global private equity market transactions in 2019 plateaued at 1.47 trillion USD versus 1.49 trillion USD in 2018. Before 2019, the amount of global private equity market transactions has grown 12% annually from 2013 to 2018.

As the primary market lacks standardization, information asymmetry, fraud, and high cost on due diligence have led to seriously fragmented ecosystems. Industry and academia have tried to change this status quo, but there is no noticeable effect. The infrastructure of the primary financial market is not so easy to optimize without solving the challenges listed.

Since Bitcoin was invented, the blockchain technology which supports it has quickly entered people's eyes [3]. One of the essential purposes of the invention of blockchain was to develop the financial industry [4]. The decentralized idea and unique features of blockchain, such as decentralization, highly transparent, enhanced security, and immutability of information, make blockchain be the most appropriated technology to along with the logic of financial markets [5].

To help advance research in primary financial market infrastructure development, we conducted an empirical study to investigate the work practice and potential challenges faced by primary market involvers. We followed a blended strategy approach that combines interviews (qualitative method) and surveys (quantitative method). In particular, we interviewed a total of 15 primary financial involvers with various backgrounds and expertise. We asked participants about their everyday work and relevant challenges faced during their responsibilities in the primary financial market during the interviews. Next, we adopted open card sorting [6] to analyze the interview results. The following categories produced by open card sorting were grouped into three groups, i.e., complex due diligence, mismatching, and difficult monitoring. After that, we performed a validation survey with 54 participants to confirm various insights from the interviews, including challenges, best practices, and desired improvements based on the interviews.

---

[1] The primary author: jliu3872@uni.sydney.edu.au
[*] The correspondent author: jliu3872@uni.sydney.edu.au

According to the interview and survey, we realized that the primary market participants cared a lot about due diligence but did not effectively avoid the complex process. Besides, the asymmetric information and lack of trust among the participants make it hard to settle. Also, it is hard to discover each other, which makes the market lack liquidity. The limitations influenced their day-by-day work, particularly for large companies.

The significant contributions of our research are as follows:

• As far as we know, this is the first in-depth research investigating primary market participants' insights on the current status of the primary financial market and blockchain technology through interviews and surveys.

• We analyze both qualitative and quantitative data and highlight potential opportunities and implications that investors, financiers, brokers, and other financial market participants can use to improve their daily work under a new infrastructure in the future.

The following sections are organized as below: In the next section, we provide background materials on primary market and blockchain. In section 3, we present our 2-parts' methodology, both quantitative and qualitative. The findings of our study are discussed in Section 4. There are some potential research directions according to our findings in Section 5. In Section 6, we discuss the threats to the validity of our research. Section 7 is the related work. The final section gives conclusions and future work.

## II. BACKGROUND

### A. Primary Financial Market

Since the 1970s, the development of information technology has vigorously promoted the popularization and application of the electronic and networked transaction settlement system in the financial market, realizing the rapid development of the multi-level securities market and the informatization of market participants. The most prominent feature of securities trading is a centralized third-party credit or information intermediary agency as a guarantee, which realizes value transfer by reducing information asymmetry and relying on traditional institutions to establish a trust mechanism, no matter in the primary or secondary market.

Because of the unique characteristics of the primary financial market, it seems impossible to have a unified exchange system in the world [7]. In practice, the following process shown in Fig. 1 is how to conduct simple private equity transactions in primary market:

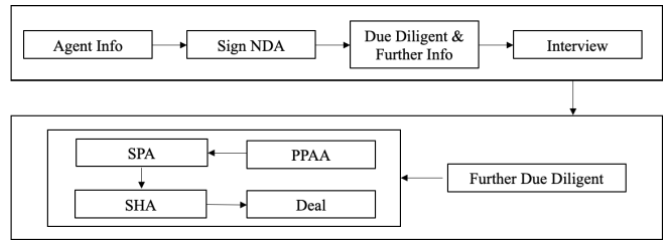

Fig. 1. Workflow of primary market.

Buyers and sellers know each other through the introduction of a broker, sometimes through a friend's introduction, but most transactions are reached through a broker, which is usually called an FA (Financial Advisor). After the buyer and seller contact, if the buyer and seller intend to continue the transaction, they will sign a non-disclosure agreement (NDA). After signing the NDA, the seller sends some of its basic information to the buyer, and the buyer will need to hire third-party service companies to validate the authenticity.

After the buyer reconfirms that this firm or project is the target it is interested in, the buyer will forward to the next step, an on-site interview. After the interview, the DD (Due Diligence) process will be launched, and if both the interview and DD meet the buyer's requirements, they will sign a TS (Term Sheet), which can be regarded as a promised investment agreement. TS is a letter of intent to invest and is not subject to legal restrictions. Finally, the buyer and seller will sign SPA (Share Purchase Agreement) and other relevant agreements to confirm the investment contract, which is the final step of the investment activity. Other relevant agreements normally include SHA (Shareholders Agreement) and PPAA (Predict Profit Allocation Agreement).

Trust, data sharing, data security, and related personalization characteristics are the main reasons for the existing problems in the primary market [8, 9, 10, 11, 12, 13, 14, 15, 16, 17, 18, 19, 20, 21, 22, 23].

### B. Blockchain

Blockchain technology is another disruptive technology after cloud computing, the Internet of Things, and big data. Blockchain can improve the efficiency of digital collaboration in the actual business process and provide a value-based interconnection infrastructure for the financial industry and trading methods inside the financial market. According to its unique characteristics of "decentralization" and "trust," proper application of blockchain technology can ensure that trust is established between multiple parties while protecting data privacy and is expected to become an essential part of the financial market infrastructure.

An essential purpose of blockchain technology innovation is to provide services for financial transactions. It is a distributed ledger for transactions with an append-only feature.

Blockchain was first proposed by Bitcoin [24]. It was initially a decentralized electronic instalment framework, thereby eliminating any external requirements related to instalment activities. The first Bitcoin blockchain combined information into a chain, coordinated by square hash values named "blockchain." As stated by [24], Bitcoin is the first form to use blockchain as the overall framework of a distributed system.

## III. METHODOLOGY

Figure 2 illustrates the overview of our methodology design, which combines the interview part and survey part. Interview part: 15 experts from the primary financial market are interviewed to get insights into the primary market. Survey part: it is used to validate the findings from the interviews. We present how we design and implement our interview and survey in the following.

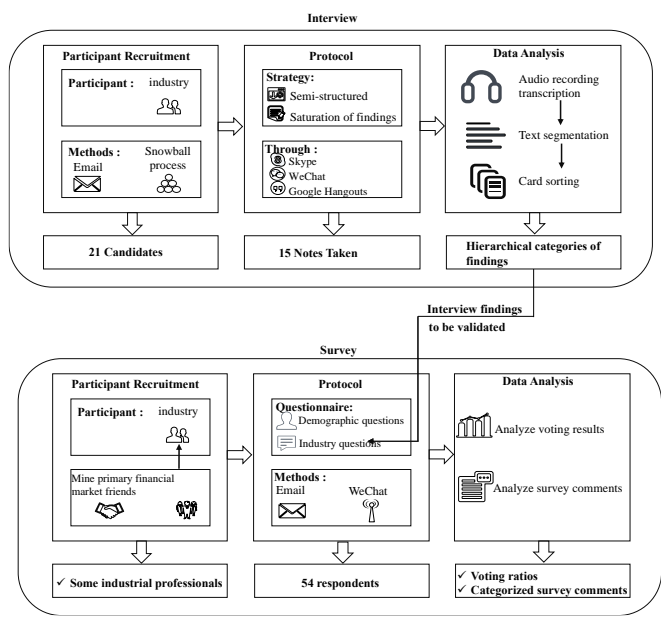

Fig. 2. Methodology design.

### A. Interview

In our research, we adopted semi-structured interviews [25]. We first exchanged our introductory information, such as respective institutions and positions, with the interviewees and then introduced our research and purpose. Next, we asked some basic qualitative questions to the interviewees. Then, we used some open questions to guide the interviewee with our directions (Questions are listed in Table 1). These open questions probed the interviewees about their perspectives on the primary financial market and applying blockchain in the primary market. Since the design of semi-structured interviews, we arranged subsequent questions to delve further into our interview participant's perspectives at a later stage. In the end, we asked that the interviewee could give whatever other significant data that we may have missed during the interviews.

| 1. | What is your role in the primary market? Please describe your main duty and goal. |
|---|---|
| 2. | What are the 3 most common pain points/difficulties in your business? |
| 3. | Let us discuss these pain points one by one: <br> 1) Regarding pain point 1, please describe in detail, can you explain what you think is causing it? <br> 2) Regarding pain point 2, please describe in detail, can you explain what you think is causing it? <br> 3) Regarding pain point 3, please describe it in detail. Can you explain what you think is causing it? <br> 4) Optional: If the respondent does not list any cost, trust or efficiency issues, we will need the following questions. What do you think of these issues in the industry? |
| 4. | In view of the above pain points, which one do you think needs to be solved urgently? Do you know or understand any innovations/solutions to these problems? |
| 5. | Have you heard of blockchain? (If not, please skip; if yes, please continue to ask). Do you think that applying it in the industry can solve the above problems? |
| 6. | If a product/technology can solve the above problems in some way, will your industry group be willing to use it? |
| 7. | Do you think you have any relevant supplements in our interview today? For example, you think it is important, but we forgot to ask. |

Table 1. Problems in primary financial market and solutions from blockchain

In this interview, a total of 15 interviewees were interviewed. Among them, 5 are at or above the partner level of equity investment institutions, 4 are from third-party service agencies, 4 are corporate financing directors or directors of the financing department, and 2 are from investment banks with solid experience. The interviewees covered the prominent participants in the primary market.

We have a total of 21 candidates this time, but due to uncontrollable reasons, such as covid-19 and timing, only 15 of them were interviewed by us in the end. During the interviews, we followed the procedure utilized in [26] and [27] to choose when to stop the meeting, i.e., halting interviews when there is enough saturation of findings. Saturation is a methodology that is widely used in qualitative research [28], [29], [30].

To ensure that our interviews are of broad-spectrum significance, our interviewees covered most types of primary market participants, such as investment institutions, financing companies, law firms, accounting firms, etc.

We made sure to interview interviewees from various backgrounds (as demonstrated in Table 2) before deciding whether saturation had been reached. In every interview, we cooperated to pose questions and take notes. After completing each interview, we would contrast their notes with past ones to check whether there are any new insights from the interview.

Because of the Covid-19, all interviews were conducted remotely via Zoom and WeChat, and the interviewers took notes. The average and standard deviation of the interview time were 35 and 30 minutes, respectively. Table 2 illustrates the basic demographics of the interviewees. According to the table, the interviewees had an average experience of 12.13 years working experience and 10.4 years in the primary financial market by the time of interviews.

| No. | Role | General Experience | Primary Market Experience |
|-----|------|--------------------|---------------------------|
| 1 | PE/VC Partner | 10 | 10 |
| 2 | Financing VP | 17 | 15 |
| 3 | PE/VC Partner | 12 | 11 |
| 4 | PE/VC Director | 11 | 6 |
| 5 | PE/VC Director | 8 | 8 |
| 6 | PE/VC Director | 9 | 9 |
| 7 | Financing VP | 12 | 10 |
| 8 | CFO | 13 | 10 |
| 9 | PE/VC Partner | 16 | 13 |
| 10 | Financing VP | 12 | 8 |
| 11 | Financing VP | 22 | 18 |
| 12 | PE/VC Director | 14 | 12 |
| 13 | PE/VC Partner | 10 | 10 |
| 14 | Financing VP | 7 | 7 |
| 15 | PE/VC Director | 9 | 9 |

Table 2. Interviewee's background.

### 1. Participant Recruitment

Two of the authors used to work in the primary financial market, and they used their connections to contact an initial group 10 candidates. We utilized a snowball process to generate another group with another 11 candidates [31], i.e., current participants refer the interview to their target participants. There are a total 21 experts agreed to take part in the interview.

### 2. Data Analysis

We used card sorting [6] to recognize the classifications from each interview. It is a particular procedure to get classifications from data [32], [33]. There are three different types: closed card sorting with predefined categories for data, open card sorting with no predefined categories, and hybrid card sorting, which combines the previous two types [34]. Considering our research is exploratory with categories (i.e., challenges of the primary market) being unknown in advance, we chose to conduct an open card sorting process to analyze these data.

In particular, after a card was made for each textual unit in the card sorting, the cards were then bunched into influential groups with a theme or topic meanings: these groups, i.e., low-level, mid-level, and high-level categories. The consequences of a remarkably open card sorting would allow us to acquire various hierarchical designs of categories. Two authors were involved in the card sorting process. Every card was identified and analyzed by them. We identified three high-level categories by using card sorting, i.e., complex due diligence, mismatching, and difficult monitoring.

## B. Survey

### 1. Design

Our questionnaire includes three demographic questions, 7 primary financial market questions, and 2 blockchain-related questions. The demographic questions are single choice and designed to understand the background and experience of participants. The primary financial market questions are designed to validate insights that we have found from our interview part. The participants who have been interviewed are not asked to respond to the survey. There are both single choices and 2 choices questions included. We regard the 2 choices questions as sorting to analyze more straightforward rather than plan sorting method. The blockchain-related questions are mainly designed for validating our idea that blockchain technology will improve the infrastructure of the primary financial market. The complete list of our survey questions can be found in Appendix 1.

### 2. Survey Respondent Recruitment and Statistics

Our potential survey participants are professional experts involved in the primary financial market, their primary role including investment, financing, brokerage, audit, etc. Since the main participants are from China and some are from Australia and the United States, we have made Chinese and English versions of the questionnaire. This survey is released through the application of Questionnaire Star. The respondents can use any mobile phone/computer/pad to respond to the survey by scanning the code.

When selecting respondents, we pay more attention to their work experience in the primary market, preferably compound working experience, to make a more efficient result. We do not use other factors such as academic qualifications to limit the selection of respondents. Because most of the participants in the primary financial market have higher academic qualifications, and all the participants we selected have master or doctoral degrees. To eliminate similar ideas from the same company or the same period of work, we control that at most two of our respondents are from the same company.

We finally selected 54 respondents. From the perspective of the participants' roles, 40.7% of them are from investment institutions, 24.07% from financing parties, 20.37% from third-party service intermediaries, and 14.81% from brokers. For general working experience, 70.37% of people have more than 4 years of work experience, and only 29.63% have less than 3 years of work experience. For primary financial market working experience, 55.56% of people have more than 4 years of primary market experience, and only 44.44% of people have less than 3 years of relevant work experience. Pie charts of broad working experience and primary financial market working experience are shown in fig. 3. It is a relatively convincing work background structure.

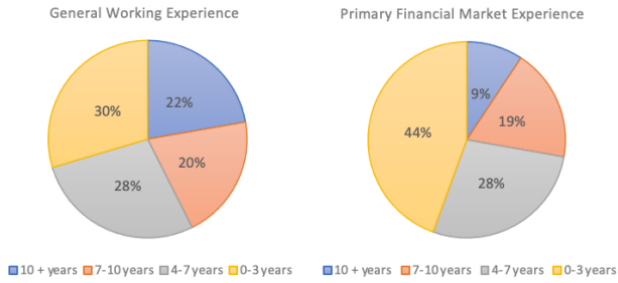

Fig. 3. Pir charts of general working experience and primary financial market experience in percentage.

*3. Data Analysis*

After we terminated the questionnaire collection, we analyzed the closed-ended questions by adopting different analysis methods. We calculated the number of votes for each answer option. Next, we counted its percentage rate for each answer option by dividing the number of votes for the option by 54 (total respondents' number).

To better understand respondents' perspectives of the primary financial market's challenges, we divided the participants into several demographic groups and compared their voting results with each other's. Based on previous studies [52] and [53], we constructed the following demographic groups:

• Respondents from the investment field (Inv)

• Respondents from the financing field (Fin)

• Respondents from third parties (Thi)

• Respondents from brokerage (Bro)

• Respondents with high general working experience (>=10 years) (GExpH)

• Respondents with low general working experience (<=3 years) (GExpL)

• Respondents with medium general working experience (>3, and <10 years) (GExpM)

• Respondents with high primary financial market related working experience (>=7 years) (FExpH)

• Respondents with low primary financial market related working experience (<=3 years) (FExpL)

• Respondents with low primary financial market related working experience (>3, and <7 years) (FExpM)

We calculated the percentage ratios of their answers to the primary financial market challenges and desired improvements collected from the interview separately for each demographic group. According to [54], we applied Fisher's exact test [55] with Bonferroni correction [56] to these numbers to make sure if one group tended to vote differently with others. Fisher's exact test reveals the frequency distribution of the variables (e.g., each option votes from each group in our research) in the analysis of contingency tables. It can determine if the observed difference between two proportions (i.e., the ratio of votes) is statistically significant. The family-wise error could be controlled by Bonferroni correction while making multiple comparisons. Section 4.4.2 shares the analysis results in detail.

## IV. FINDINGS

In this part, firstly, we present our findings for each category (totally three categories) that were identified by utilizing open card sorting on the interview data. There are subcategories for each category, and we select some of the most effective content and analyze some statistics according to our survey feedback to highlight the generality of these findings. Next, we introduced the voting results of each population group in response to these challenges and the potential solutions mentioned by the interviewees and the related significance tests of these results. Finally, we briefly conclude our interview and survey results.

### A. Complex Due Diligence

The due diligence process of primary market transactions is mainly carried out through written review, on-site inspections, public channel searches, interviews, and entrusted third-party investment. As investors continue to pay more attention to issues such as consistency of interest and information transparency, the level of detail in their preliminary due diligence process has also increased. Although detailed and meticulous due diligence meets investors' specific due diligence requirements for companies to some extent, it often takes too long, reduces transaction efficiency, and increases transaction costs. On the other hand, due to the lack of a standardized due diligence checklist, some investors do not know the extent of their due diligence on the company.

### B. Mismatching

The primary market is dominated by "over-the-counter transactions". The two parties may negotiate between the two parties or through the "matching" of third-party service agencies to complete the transaction. In this process, there is asymmetric and redundant information in the transaction, making it difficult for both parties to match the transaction accurately, and the transaction time is lengthened indefinitely. At the same time, some trading platforms lack the guarantee of credibility, and the ability to label information is weak. It is not easy to form a mature trading loop in the primary market. This also leads to low matching efficiency between the two parties in the primary market, and it is not easy to achieve transactions quickly.

### C. Difficult Monitoring

Post-investment monitoring means that investors need to understand the company's trends in time, understand the direction of capital use, identify problems in the company's development process, help companies carry out standardized management, and to a certain extent restrict and deter the company. It is necessary to grasp the appropriate degree, which can meet your own needs and do not make the invested company feel too troublesome and controlled in the monitoring step. At present, a considerable part of the post-investment work in the primary market is in a state of groping. Post-investment work is relatively casual, lacks standardized procedures, and lacks assessment standards. The most direct consequence is that post-investment personnel do not know what to do and cause laxity. Investors seem to have paid human resources and financial resources, but the actual monitoring effect is unsatisfactory.

### D. Survey Result

#### 1. Interview

##### a) Primary market pain points and difficult issues

Primary market transactions' pain points and difficulties run through all the links of "funding, investment, monitor and withdrawal". Funding is to raise funds. People who need to maintain or increase their assets will invest the money in investment institutions for management. Investment is to find a good company or project to invest in equity or debt. The monitoring is to track and manage invested projects and conduct follow-up reviews. Withdrawal is the process of finally selling the investment product and cashing out.

Among the 15 respondents, 14 respondents (93.3%) believed that the current due diligence procedures for securities issuance are too complicated, time-consuming, and costly, and they are on the premise of ensuring the quality of issuance. The issuance efficiency will be affected to a certain extent under complex due diligence. 10 respondents (66.67%) believe that it is difficult to match the needs of investment institutions with the needs of invested companies. For the investment process of equity investment, 7 (46.67%) believe that the drawer agreement is difficult to implement. Regarding the follow-up management of equity investment, 11 respondents (73.3%) believe that the effective implementation of post-investment management is facing specific difficulties and that there are challenges in fully obtaining various types of information on invested companies. At the same time, due to the imperfect exit mechanism, lack of professional intermediaries, and insufficient attention to exit management in the primary market, it is difficult for the primary market to exit equity.

##### b) Causes of the issues in primary market

All respondents agreed that the above-mentioned pain points and difficulties in the current primary market are rooted in two aspects: lack of trust (100%) and difficulty in ensuring the authenticity of data (100%).

##### c) Issues that need to be solved in primary market

Faced with the current pain points and difficulties in the primary market, respondents believe that the complexity of the due diligence process (93.3%) and post-investment management and exit (66.67%) should be resolved first. In addition, 58.65% of the respondents believe that improving the matching of the needs of investment institutions and invested companies is also an urgent problem that needs to be resolved.

##### d) Can the blockchain be used to solve the pain points and difficult issues of the primary market?

Although all the interviewees have varying degrees of understanding of blockchain technology, there are still some disagreements on whether blockchain technology can solve the pain points and difficulties of the primary market. 86.67% of the respondents have a positive attitude towards the application of blockchain technology, believing that based on the advantages of blockchain technology, the application of this technology can solve some of the pain points and difficulties. Furthermore, for the specific application scenarios of blockchain technology in the primary market, the vast majority of interviewees did not have a deeper understanding and knowledge, and only one person was familiar with the application scenarios of blockchain technology and believed that part of the trust and process problems could be solved through the characteristics of the blockchain.

##### e) The acceptance of blockchain products in the primary market

100% of the interviewees believe that, on the premise that the transaction cost has not increased significantly if the products based on blockchain technology can better solve the pain points and difficulties of the primary market, they will choose to use related products.

#### 2. Survey

Table 3 lists 20 challenges and 7 desired improvements mentioned by interviewees in the above sections. C1 to C4 were significant challenges of the pre-investment stage. C5 to C16 were significant challenges during the investment stage, and C17 to C20 were significant challenges of the post-investment stage. I21 to I24 were desired internal optimization in the primary financial market environment in the future. I25 to I27 were expected external support in the future. The last column is the ratio (percentage) of respondents who voted for the second column challenges.

| ID | Challenges/Desired Improvements | Votes (Percentage) |
|---|---|---|
| *Pre-investment Stage* | | |
| C1 | Looking for investment channels and reliable partners | 47 (87.04%) |
| C2 | Methods and experience of screening companies | 24 (44.44%) |
| C3 | Forecast the development space prospects of the industry | 21 (38.89%) |
| C4 | Investment risk analysis | 16 (29.63%) |
| *Investment Stage* | | |
| C5 | Due diligence | 43 (79.63%) |
| C6 | Evaluation of matching needs between investment and financing parties | 43 (79.63%) |
| C7 | The authenticity of the information provided by the target company | 40 (74.07%) |
| C8 | Monitoring | 33 (61.11%) |
| C9 | Fair and objective industry/company information sources | 30 (55.56%) |
| C10 | Fundraising | 26 (48.15%) |
| C11 | Find high-quality companies as investment targets | 23 (42.59%) |
| C12 | Investment decision and execution | 21 (38.89%) |
| C13 | Performance evaluation | 20 (37.04%) |
| C14 | Realization of investment income | 18 (33.33%) |
| C15 | Professional and authoritative judgments on the development trend of the target | 15 (27.78%) |

| C16 | Information system support | 12 (22.22%) |
|---|---|---|
| | *Post-investment Stage* | |
| C17 | Exit mechanism | 38 (70.37%) |
| C18 | Information transparency | 31 (57.41%) |
| C19 | Post-investment risk management | 27 (50.00%) |
| C20 | Participation in major decisions of the company | 12 (22.22%) |
| | *Desired Internal Optimization* | |
| I21 | Integrate internal and external data to better analyse and judge service investment | 21 (38.89%) |
| I22 | Use new technology to improve the efficiency of investment process | 16 (29.63%) |
| I23 | Use new technologies to match investment needs more intelligently | 11 (20.37%) |
| I24 | Understand the application mode and applicability of new technologies in investment | 6 (11.11%) |
| | *Desired External Support* | |
| I25 | Professional resource support, providing industry/company evaluation and judgment information | 21 (38.89%) |
| I26 | Docking with outstanding target companies and tapping investment opportunities | 20 (37.04%) |
| I27 | Use third-party service organizations to apply and develop new technologies | 13  24.07%) |

Table3. Votes of challenges and desired improvements from interview.

After analyzing the overall voting result of individual challenges and desired improvements, we will analyze them by different demographic groups. Table 4 illustrates the detailed results of the voting.

| ID | Inv | Fin | Thi | Bro | GExpH | GExpL | GExpM | FExpH | FExpL | FExpM |
|---|---|---|---|---|---|---|---|---|---|---|
| Total | 22 | 13 | 11 | 8 | 12 | 16 | 26 | 15 | 24 | 15 |
| C1 | 81.82 | 100.00 | 81.82 | 87.50 | 91.67 | 75.00 | 92.30 | 93.35 | 79.17 | 93.33 |
| C2 | 40.91 | 38.46 | 45.45 | 62.50 | 50.00 | 37.50 | 46.15 | 46.66 | 37.50 | 53.33 |
| C3 | 27.27 | 53.85 | 45.45 | 37.50 | 41.67 | 50.00 | 30.76 | 26.68 | 50.00 | 33.33 |
| C4 | 50.00 | 7.96 | 27.27 | 12.50 | 16.67 | 37.50 | 30.76 | 33.34 | 33.33 | 20.00 |
| C5 | 77.25 | 92.31 | 63.64 | 87.50 | 75.00 | 75.00 | 84.62 | 86.67 | 79.17 | 73.33 |
| C6 | 86.36 | 61.54 | 90.91 | 75.00 | 75.00 | 68.75 | 88.46 | 86.67 | 70.83 | 86.67 |
| C7 | 68.18 | 84.62 | 72.73 | 75.00 | 83.33 | 68.75 | 73.08 | 80.00 | 70.83 | 73.33 |
| C8 | 63.64 | 46.15 | 63.64 | 75.00 | 66.67 | 62.50 | 57.69 | 66.67 | 66.67 | 46.67 |
| C9 | 59.09 | 61.54 | 54.55 | 37.50 | 50.00 | 37.50 | 57.69 | 66.67 | 41.67 | 66.67 |
| C10 | 54.55 | 61.54 | 18.18 | 50.00 | 58.33 | 50.00 | 42.31 | 46.67 | 54.17 | 40.00 |
| C11 | 50.00 | 23.08 | 45.45 | 50.00 | 41.67 | 56.25 | 42.31 | 33.33 | 50.00 | 40.00 |
| C12 | 31.82 | 30.77 | 54.55 | 50.00 | 8.33 | 37.50 | 53.85 | 33.33 | 41.67 | 40.00 |
| C13 | 36.36 | 46.15 | 45.45 | 12.50 | 41.67 | 37.50 | 34.62 | 26.67 | 66.67 | 53.33 |
| C14 | 27.27 | 46.15 | 36.36 | 25.00 | 58.33 | 43.75 | 15.38 | 33.33 | 33.33 | 33.33 |
| C15 | 22.73 | 30.77 | 27.27 | 37.50 | 25.00 | 37.50 | 23.08 | 20.00 | 37.50 | 20.00 |
| C16 | 22.73 | 15.38 | 27.27 | 25.00 | 16.67 | 25.00 | 23.08 | 20.00 | 20.83 | 26.67 |
| C17 | 81.82 | 61.54 | 45.45 | 87.50 | 83.33 | 75.00 | 53.85 | 80.00 | 75.00 | 53.33 |
| C18 | 50.00 | 53.85 | 81.82 | 50.00 | 41.67 | 56.25 | 38.46 | 53.33 | 54.17 | 66.67 |
| C19 | 50.00 | 46.15 | 54.55 | 50.00 | 50.00 | 50.00 | 65.38 | 46.67 | 54.17 | 46.67 |
| C20 | 18.18 | 38.46 | 18.18 | 12.50 | 25.00 | 18.75 | 42.31 | 20.00 | 16.67 | 33.33 |
| I21 | 54.55 | 30.77 | 27.27 | 25.00 | 25.00 | 37.50 | 46.15 | 40.00 | 37.50 | 40.00 |
| I22 | 22.73 | 38.46 | 36.36 | 25.00 | 50.00 | 18.75 | 26.92 | 33.33 | 29.17 | 26.67 |
| I23 | 9.09 | 23.08 | 27.27 | 37.50 | 16.67 | 25.00 | 46.15 | 20.00 | 16.67 | 26.67 |
| I24 | 13.64 | 7.69 | 9.09 | 12.50 | 8.33 | 18.75 | 38.46 | 6.67 | 16.67 | 6.67 |
| I25 | 45.45 | 23.08 | 36.36 | 50.00 | 33.33 | 43.75 | 38.46 | 46.67 | 33.33 | 40.00 |
| I26 | 13.64 | 69.23 | 45.45 | 37.50 | 50.00 | 43.75 | 26.92 | 40.00 | 41.67 | 26.67 |
| I27 | 40.91 | 7.69 | 18.18 | 12.50 | 16.67 | 12.50 | 34.62 | 13.33 | 25.00 | 33.33 |

Table4. Voting results of different groups towards 20 challenges and 7 desire improvements highlighted by questionaries. The Total row illustrates the number of respondents in each group. The rows C1 to I27 represent the percentages (%) of respondents from each group.

From Table 4, we could observe that the voting results varied from demographic groups. For example, for C17, FexpH and FExpL were 80% and 75%, while the ratio was only 53.33% for group FExpM. Another example, for I22, the ratios of GexpH, GexpL, and GExpM were 50%，18.75%, and 26.92%, respectively. To check whether the observed ratio differences are statistically significant, for each challenge/desired improvement, we applied Fisher's exact test with Bonferroni correction on three sets of demographic groups, i.e., groups with different roles (Inv vs. Fin vs. Thi vs.Bro), groups with different general working experience (GExpH vs. GExpL vs. GExpM), and groups with different primary financial market-related working experience (FExpH vs.FExpL vs.FExpM).

After conducting 270 (10 group pairs ⇾ 27 challenges/improvements), Fisher's exact tests with Bonferroni corrections found that three tests showed that the relevant difference is statistically significant. It is Inv vs. Fin L on I26 (p-value=0.002<0.05/6 after Bonferroni correction).

Based on the testing results, we can say with some certainty that: respondents from the financing field (Fin) are significantly more likely to rate I26 (Docking with outstanding target companies and tapping investment opportunities) as a significant desired external support than those respondents from investment field (Inv) (69.23% vs. 13.64%).

*E. Summary of Results*

From the analysis of the interview and survey, we could find the following:

It is urgent to solve problems with complex due diligence processes, mismatching, and difficulty monitoring. Traditional solutions are challenging to eliminate the cost increase and efficiency loss caused by these problems. In the future, you can consider choosing emerging technologies, such as blockchain, to solve these problems in the primary market. The current traditional solution is to reduce the friction of information/trust issues by introducing financial agency, proving lack of efficiency.

V. FUTURE DIRECTIONS

Blockchain technology has the potential to bring the infrastructure of the primary financial market into the next generation.

The blockchain has an available feature. Except that the private information of each transaction party is encrypted, market participants can query and obtain data on the blockchain through a public interface. Therefore, the information transparency of the entire system is extremely high, and it is more convenient for transaction parties to obtain information, which is conducive to reducing the information asymmetry of both parties in the market, improving the matching efficiency of both parties, and faster settlement.

The use of blockchain technology can reduce the reduction in transaction efficiency and the increased transaction costs caused by the excessively long due diligence process. Using the consensus of blockchain technology, tamper-proof, traceable features, and the automaticity of timely update of node information, all aspects of information involved in the transaction can be recorded on the blockchain. Therefore, investors can clearly understand and view the investment and, based on ensuring the credibility of assets, minimize the complicated process involved in the due diligence process.

The block-based intelligent contract technology can help non-standard off-site contracts be presented in executable code form, making it easier to realize automated transactions of non-standardized contracts. Each market participant in the blockchain has a complete transaction record, and the transaction assets are anchored on the blockchain. The conclusion of the contract becomes flat and automatically executed, which improves the ability of both parties to perform the contract and reduces the transaction counterparty risk.

For post-transaction events, like derivatives transactions and clearing, etc., rely on blockchain technology to manage the DLT network, monitor related assets, and redesign and optimize related processes, improving market transparency and enhancing the efficiency of derivatives transaction management and reduce transaction costs. Benben.

The peer-to-peer transaction method based on blockchain decentralization technology, under the action of the consensus mechanism, makes consensus and mutual trust automatically reached when both parties of the transaction exchange data, which not only ensures information security but also effectively improves efficiency and reduces transaction cost.

## VI. THREATS TO VALIDITY

### A. Internal Validity

In this research, our survey questions are designed based on the conclusion of the interview. However, we may sometimes misunderstand or fail to fully understand the intention of the interviewees. In order to eliminate this threat, we slowed down as much as possible during the interview and confirmed the content once we did not understand clearly at any time. The card sort step is handled by two authors together, so there is the possibility of mistakes, but we have tried our best to avoid making mistakes.

Interviews and surveys also have the possibility of interviewees providing dishonest answers for any reason. To reduce this bias, we have made the following efforts in the survey: 1) In the letter of invitation, we stated that we would not publish and try our best to avoid leaking personal information (if provided) and it is entirely confidential; 2) Our survey is anonymous, and it is guaranteed that we will not track the participants by answering the questionnaire content. If the interviewee wants to provide contact or other personal information, it can be added voluntarily. Based on [35], confidentiality and anonymity might help to obtain honest answers from interviewees.

In addition, based on the following suggestion [36], that is to provide services to prospective respondents in an appropriate language. Besides the English version of the survey, we also translated the survey into Chinese. The Chinese version of the survey can help Chinese respondents more easily understand the questions from our questionnaire. Again, we may also draw wrong analysis based on the answers of the survey. To avoid this threat, we chose to read their answers as carefully as possible.

### B. External Validity

According to our interview strategy [37, 38], we only interviewed 15 interviewees because we believe that the saturation has reached. We admit that our number of interviews is not very large, and the strategy is saturated and cannot fully represent all situations, but our interviewees have been reset to cover all essential primary market roles.

Considering that some interviewers will have some ideas or opinions that we might miss or have new and meaningful ideas after being interviewed by us, all our interviewees are experienced. Their rich work experience will increase the completeness of the answers to open questions.

To validate the results of our interview, we surveyed 54 participants. Since most of the survey respondents are from China, Australia, and the United States, we cannot guarantee that our survey results apply to the world. It is anonymous and does not require identity verification, so there is no guarantee that the interviewees we are looking for including all participants in the entire primary financial market. To further improve the generalizability of the research results, we encourage other scholars to replicate our research with a more extensive group of participants in the future.

## VII. RELATED WORK

This section highlights related work on applying blockchain technology into the primary financial market, including proposed systems to help the primary ecosystem, core technologies of blockchain, and core advantages and applications of blockchain.

### A. Proposed Primary Financial Market Systems

Blockchain is continuously innovated and expanded on top of Bitcoin's infrastructure. At present, blockchain can be divided into the public chain, alliance chain, and private chain according to the access mechanism of nodes and the degree of decentralization. Blockchain technology has gone through the Bitcoin era of blockchain 1.0 and the blockchain 2.0 era represented by the alliance chain. At present, blockchain technology has transitioned to the blockchain 3.0 era represented by EOS. In terms of technical application, according to different actual application scenarios and design concepts, current blockchain projects are heterogeneous blockchains developed using different technical frameworks.

At present, there are relatively few articles talking about the application of blockchain technology in the primary financial market in practice compared with the secondary market. However, the relevant issues still attracted a small group of scholars' attentions.

In terms of securities issuance, [40] points out that replacing the securitization third party in the primary market with a blockchain-based system can more accurately track process details, reduce costs, increase issuance speed, increase transparency and liquidity. [41] believes that the fraud issues of information asymmetry in the primary market can be reduced by adopting blockchain technology in the financial market infrastructure. It proposes a Linked Data-based model, which provides both data verification and tamper-proof functions to prevent collusion and increase trust in financial markets effectively.

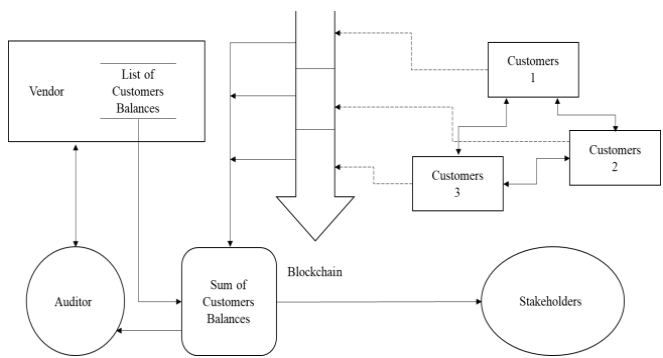

Fig. 4. Public and private access for accounting information system [42].

Authorized participants may get all information on the blockchain immediately, and the information accessed by all parties is consistent, which saves a lot of due diligence expenses and probable errors in the copying process, according to [42]. It proposed a prototype in Figure 4 for an accounting information system with public and private access to record the entity's data of access or update for the information. It will improve the accuracy of financial reporting information provided to shareholders and other approved persons. [43] demonstrates that due diligence data is on-chain, open, and transparent, significantly increasing information transparency, reducing information asymmetry, and lowering fraud risks. It presented the ORM model in Figure 5, which includes a copy of the data before entering the database and a blockchain-verified mirror file. The blockchain will genuinely update the verified data given to the database after relaying it back to the database, preventing the original information from being tampered with.

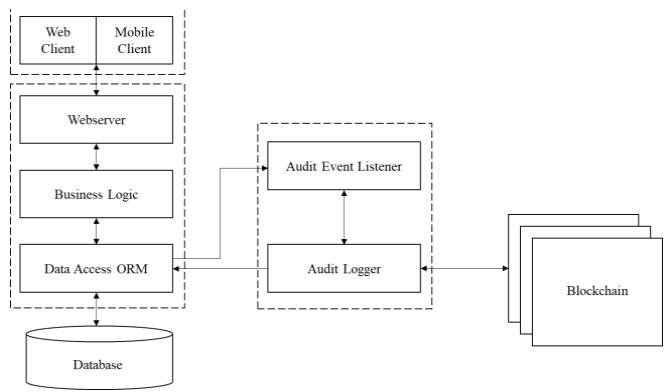

Fig. 5. ORM model [43].

Scholars are also debating how blockchain technology may help save expenses and increase efficiency. [44] claims that using blockchain to promote corporate governance and legal frameworks may promote corporate transparency and efficiency when the record is established, and that blockchain technology has enhanced automation. [45] shown that when pre-set criteria are satisfied, equity transactions may be automatically completed by putting smart contract terms into an automated programming language. More specifically, it employs machine learning to assess the desire and ability of borrowers, and it leverages a blockchain-based system to predict the borrower's future willingness instantaneously.

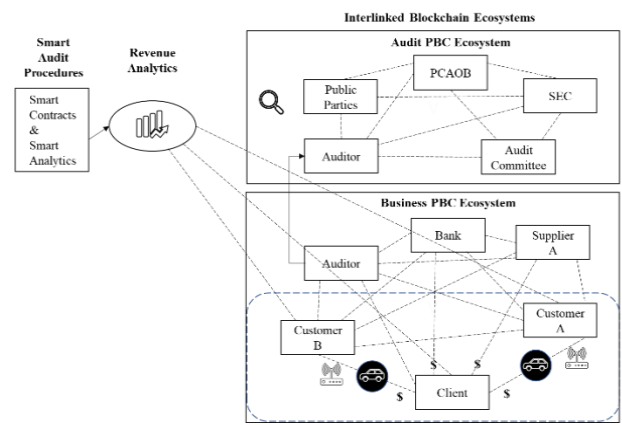

Fig. 6. Interlinked Blockchain Ecosystems [49].

Furthermore, several publications are based on a governance standpoint, believing that apps aid departmental monitoring. [46] outlines why blockchain technology can replace finance agencies and discusses the history of employing blockchain in the primary market at an early level. [47] describes how the supervision department identifies the nodes on the chain and acquires the public key to monitor the fundamental data in real-time. At the same time, owing to the blockchain's infallibility, the information's validity and traceability are ensured. [48] show that auditing and supervision might no longer be limited to sampling but will collect and process all data via a blockchain network that is maintained and shared. In contrast to the previous architecture (Fig. 6), the auditor can evaluate the data inside before going outside [49]. After the report is completed, the auditor will place it in the external ecosystem to prevent external institutions and customers from altering the data regularly. It eliminates data transmission manipulation and leakage while increasing data authenticity.

[50] proposed a paradigm for considering modifications to an existing Blockchain, which is depicted in Figure 7. When someone changes to financial or commercial data, the information is sent to all blockchain participants. The operator will have a tough time removing the signs of their change. Data exchange assists the system inefficiently in preventing property fraud.

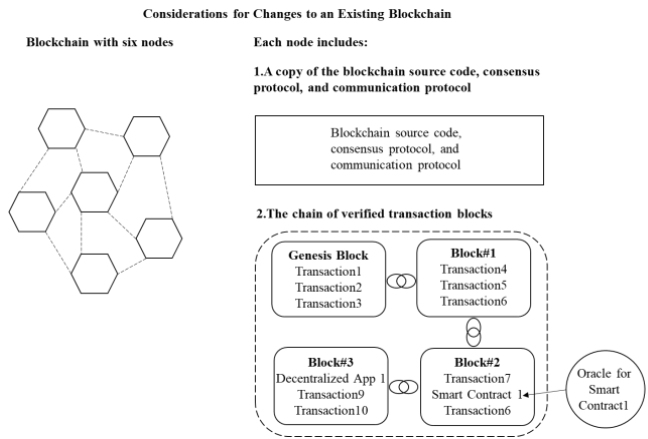

Fig. 7. Considerations for Changes to an Existing Blockchain [50].

Although the deployment of blockchain in the primary market is technically viable, the present progress in practice is modest. It is mainly owing to regulatory constraints, and no consistent norm exists [51]. Some current regulatory prohibitions will be lifted because of the government's acknowledgment of blockchain and the technological progress of distributed systems in the future.

## VIII. Conclusion and Future Work

Based on our research of combining qualitative and quantitative methods, we concluded that the most concerning issues in the primary financial market are complex due diligence, mismatch, and difficult monitoring. When the current challenges are combined with the advantages of blockchain technology, we find that blockchain may be a suitable technology that can solve the above challenges in primary financial markets. In the future, we will focus on our findings, along with blockchain and other technologies, to build a better infrastructure for primary financial market participants.

Incentives, governance, and consensus created by mathematical laws, making it possible to realize asset transactions without resorting to third-party financial intermediaries. This is having an impact on the underlying financial infrastructure. With the support of blockchain technology, securities investors and issuers can realize direct securities transactions and realize cross-time allocation without financial intermediary credit endorsement.